\documentclass[fleqn,twoside]{article}
\usepackage{espcrc2}
\usepackage{graphicx}
\usepackage[figuresright]{rotating}

\newcommand{\AmS}{{\protect\the\textfont2
  A\kern-.1667em\lower.5ex\hbox{M}\kern-.125emS}}

\title{Glueball Properties at Finite Temperature}

\author{Noriyoshi Ishii\address{Radiation Laboratory,
	The Institute of Physical and Chemical Research (RIKEN),\\
	2-1 Hirosawa, Wako, Saitama 351-0198, JAPAN}%
	\thanks{The  lattice  calculations  have  been  performed
	on NEC-SX5  at  Osaka University.},
	Hideo Suganuma\address{Tokyo Institute of Technology,\\
	2-12-1 Ohkayama, Meguro, Tokyo 152-8552, JAPAN},
	Hideo Matsufuru\address{Yukawa Institute for Theoretical Physics, Kyoto University,\\
	Kitashirakawa-Oiwake, Sakyo, Kyoto 606-8502, JAPAN}
}

\newcommand{\Eq}[1]{Eq.({\ref{#1}})}
\newcommand{\Fig}[1]{Fig.\protect\ref{#1}}
\newcommand{\Table}[1]{Table~\protect\ref{#1}}

\newcommand{\Ndf}{N_{\rm{DF}}}
\newcommand{\Nsmear}{N_{\rm{smear}}}

\newcommand{\Tate}{\rule{0cm}{1.1em}}
\newlength{\Tatescale} 

\newcommand{\Hs}{\hspace*{1em}}

\newcommand{\Bar}[1]{\overline{#1}}
\newlength{\figwidth} 
\begin{document}

\begin{abstract}
We study the glueball properties at finite temperature below $T_c$ using
SU(3)   anisotropic  quenched   lattice  QCD   with   $\beta=6.25$,  the
renormalized anisotropy  $\gamma =  a_s/a_t = 4$  and $20^3  \times N_t$
($N_t  = 35,  36,  37, 38,  40, 43,  45,  50, 72$).   From the  temporal
correlation  analysis with  the  smearing method,  we  observe the  mass
reduction of  about 20  \% for the  lowest $0^{++}$ glueball  as $m_{\rm
G}(T) =  1.25 \pm 0.1$ GeV  for $0.8 T_c <  T < T_c$  in comparison with
$m_{\rm G} \simeq 1.5 \sim 1.7$ GeV at $T\sim 0$.
\end{abstract}
\maketitle

\section{INTRODUCTION}
Finite temperature  QCD, including the quark  gluon plasma(QGP) physics,
is one of the most interesting  subjects.  Even in the hadronic phase, a
lot of effective models suggest  the changes in the hadronic properties.
However,  lattice QCD  studies of  thermal hadron  properties  have been
inadequate  until  quite recent.   This  is  due  to the  difficulty  in
measuring hadronic two-point correlators at finite temperature.  At high
temperature,  due  to the  shrink  in  the  temporal extention  and  the
consequent decrease in the number of data, the mass measurement from the
temporal  correlations  is difficult.   Hence,  finite temperature  mass
shifts  have  been  studied  through the  spatial  correlations,  which,
however, is  afflicted with the  mixture of large  Matsubara frequencies
\cite{detar}.  Recently,  the use of anisotropic  lattice is established
\cite{klassen}, which has the  finer temporal lattice spacing $a_t$ than
the spatial one  $a_s$.  In this way, available  data increases, and the
accurate  mass  measurement  directly  from  the  temporal  correlations
becomes  possible  \cite{taro,australia,ishii1}.   Here, we  report  the
thermal properties of the $0^{++}$  glueball based on the quenched SU(3)
anisotropic lattice QCD.
\section{SMEARING METHOD}
We consider the glueball correlator as
$
	G(t)
=
	\left\langle O(t) O(0) \right\rangle
$,
$
\displaystyle
	O(t)
=
	\sum_{\vec x}
	\left(\Tate
		\tilde O(\vec  x,t) - \langle    \tilde O \rangle
	\right)
$
where  $\tilde  O(\vec x,t)  =  \mbox{Re}\mbox{Tr}(  P_{12}(\vec x,t)  +
P_{23}(\vec x,t) + P_{31}(\vec x,t))$ denotes an interpolating field for
the $0^{++}$  glueball.  $P_{ij}(\vec x,t) \in {\rm  SU(3)}$ denotes the
plaquette  operator  in  the   $i$-$j$-plain.   We  apply  the  spectral
representation to $G(t)$ to have
$
	G(t)/G(0)
=
	\sum
	C_n e^{-E_n t},
$
where  $C_n =  |\langle n  | O  | 0  \rangle|^2/\sum |\langle  k| O  | 0
\rangle|^2$  with $E_n$  being the  energy of  the $n$-th  excited state
$|n\rangle$.   Here,  $|0\rangle$  denotes  the vacuum  and  $|1\rangle$
denotes the ground state glueball.  $C_n$ is non-negative with $\sum C_n
=  1$. It is  known that  $G(t)$ receives  large contributions  from the
excited  states  and the  ground  state  contribution  is small.   As  a
consequence,  the extracted  mass always  behaves much  heaver  than the
ground-state  mass.   The problem  originates  from  the  fact that  the
``size'' of the  ordinary plaquette operator is $O(a_s)$,  which is much
smaller than the physical size  $R$ of the glueball.  Hence, the problem
can be  resolved by  generating physically extended  glueball operators,
which  is achieved  by the  smearing method  \cite{ape}.   This extended
operator is  refered to  as the smeared  operator, which is  obtained by
replacing the ordinary link variables $U_i(s)$ in the plaquette operator
$P_{ij}(s)$   by  the   fat  link   variables  $\Bar{U}_i(s)   \in  {\rm
SU(3)}$. $\Bar{U}_i(s)$ is defined so  as to maximize ${\rm Re} {\rm Tr}
\left(  \Bar{U}^{\dagger}_i(s)  V_i(s)  \right)$,  where  $\displaystyle
V_i(s) \equiv \alpha U_i(s) + \sum_{j\neq i,\pm} U_{\pm j}(s) U_{i}(s\pm
\hat  j)  U^{\dagger}_{\pm j}(s  +  \hat  i)$  with $U_{-\mu}(s)  \equiv
U^{\dagger}_{\mu}(s -  \hat{\mu})$.  $\alpha$ is a  real parameter.  The
summation  involves only the  spatial directions  to avoid  the temporal
nonlocality.    Note   that   $\Bar{U}_i(s)$   holds  the   same   gauge
transformation properties as $U_i(s)$.  We refer to the fat link defined
in this way  as the first fat link $U^{(1)}_i(s)$.   The $n$-th fat link
$U^{(n)}_i(s)$   is   defined   iteratively  by   $U^{(n)}_i(s)   \equiv
\Bar{U}^{(n-1)}_i(s)$ starting  from $U^{(1)}_i(s) \equiv \Bar{U}_i(s)$.
The plaquette operator constructed  with $U^{(n)}_i(s)$ is refered to as
the $n$-th smeared operator.
We next consider  the size of the $n$-th smeared  operator. By using the
linearization and  the continuum approximation, we  obtain the diffution
equation as
\begin{equation}
	{\partial\over \partial n}
	\phi_i(n; \vec x)
=
	D
	\triangle
	\phi_i(n; \vec x),
\Hs
	D \equiv {a_s^2 \over \alpha + 4}.
\end{equation}
$\phi_i(n;  \vec  x)$ describes  the  distribution  of  the gluon  field
$A_i(\vec x,t)$ in the $n$-th  smeared plaquette.  In the $n$-th smeared
plaquette  located at  the origin  $\vec x  = \vec  0$, the  gauge field
$A_i(\vec x,t)$ is distributed in the Gaussian form as
$\displaystyle e^{-\vec x^2/(4Dn)}/(4\pi D n)^{3/2}$.
%
%
%
Hence the size of the operator is estimated as
\begin{equation}
	R
\equiv
	\sqrt{\langle \vec x^2\rangle}
=
	\sqrt{6 Dn}
=
	a_s\sqrt{ 6n \over \alpha + 4}.
\label{size}
\end{equation}
The original aim of the smearing method is the accurate mass measurement
by maximizing the ground-state overlap.  However, it can be also used to
give a rough estimate of the  physical glueball size.  In fact, once the
maximum overlap  is achieved  with some $n$  and $\alpha$,  the glueball
size can be estimated with \Eq{size}.
\section{SU(3) LATTICE QCD RESULT}
\begin{figure}
\includegraphics[width=\figwidth]{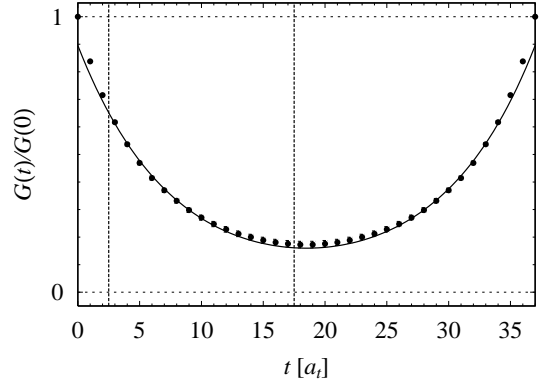}
\caption{ The  $0^{++}$ glueball  correlator $G(t)/G(0)$ for  $\Nsmear =
40$  at $T  = 245$  MeV ($N_t=37$).  The statistical  errors  are hidden
within the  symbols. The solid  line denotes the best  single hyperbolic
cosine  fit  performed in  the  interval  indicated  by vertical  dotted
lines.}
\label{correlator-37}
\end{figure}
We use the SU(3) anisotropic lattice plaquette action as
$\displaystyle
	S_G
=
	{\beta \over N_c}
	{1 \over \xi_0}
	\sum_{s, i<j \le 3}
	{\rm Re}
	{\rm Tr}
	( 1 - P_{ij}(s) )
	+
	{\beta \over N_c}
	\xi_0
	\sum_{s, i \le 3}
	{\rm Re}
	{\rm Tr}
	( 1 - P_{i4}(s) )$
with  the plaquette operator  $P_{\mu\nu}(s)$ in  the $\mu$-$\nu$-plane.
The lattice  parameter and  the bare anisotropy  parameter are  fixed as
$\beta  =6.25$,  $\xi_0   =  3.2552$,  respectively.   These  parameters
reproduce the  spatial lattice spacing as  $a_s^{-1}=2.272(16)$ GeV, and
the  temporal   one  as  $a_t^{-1}=9.088(64)$   GeV.   The  renormalized
anisotropy is $\gamma \equiv a_s/a_t=4$ \cite{klassen}.  Here, the scale
unit   is   determined   by    reproducing   the   string   tension   as
$\sqrt{\sigma}=427$ MeV from the  on-axis data of the static inter-quark
potential.  The  pseudo-heat-bath algorithm is used to  update the gauge
field configurations on the lattice  of the sizes $20^3 \times N_t$ with
various $N_t$ listed in \Table{table}.  For each temperature, we pick up
gauge  field configurations  every  100 sweeps  for measurements,  after
skipping more than 20,000 sweeps  of the thermalization.  The numbers of
gauge  configurations  used  in   our  calculations  are  summarized  in
\Table{table}.  From  the analysis  of the Polyakov  loop, we  find $T_c
\simeq 260$ MeV.  To enhance the ground-state contribution, we adopt the
smearing method with $\alpha=2.1$.
\begin{table*}
\caption{The lattice QCD result for the lowest $0^{++}$ glueball mass at
finite temperature. The temporal lattice points $N_t$, the corresponding
temperature $T$, the lowest $0^{++}$  glueball mass $m_{\rm G} (T)$, the
maximal value  of the  ground-state overlap $C^{\rm  max}$, uncorrelated
$\chi^2/\Ndf$, the best smearing  number $N_{\rm smear}^{\rm best}$, the
smearing  window  $N_{\rm  smear}^{\rm  window}$, the  number  of  gauge
configurations $N_{\rm  conf}$ and the  glueball size $R  \simeq \langle
\sqrt{\vec x^2} \rangle$ are listed.}
\label{table}
\begin{tabular}{ccccccccc}
\hline
$N_t$ &
$T$ [MeV] &
$m_G$ [GeV] &
$C^{\rm max}$ &
$\chi^2/\Ndf$ &
$N_{\rm smear}^{\rm best}$ &
$N_{\rm smear}^{\rm window}$ &
$N_{\rm conf}$ &
$R$ [fm]\\
\hline
72 & 125 & $1.54(1)$ & $0.981(6)$ & 0.23 &  41& $32 \sim 51$ & 1284 & $0.49 \sim 0.61$
\\
50 & 182 & $1.43(2)$ & $0.967(8)$ & 0.02 &  45& $36 \sim 55$ & 1051 & $0.52 \sim 0.64$
\\
45 & 202 & $1.48(2)$ & $0.971(8)$ & 0.08 &  37& $30 \sim 46$ & 1382 & $0.47 \sim 0.58$
\\
43 & 210 & $1.30(2)$ & $0.945(8)$ & 0.35 &  40& $32 \sim 52$ & 1197 & $0.49 \sim 0.62$
\\
40 & 227 & $1.29(2)$ & $0.935(6)$ & 1.22 &  40& $32 \sim 51$ & 2021 & $0.49 \sim 0.61$
\\
38 & 239 & $1.36(2)$ & $0.935(7)$ & 1.00 &  38& $30 \sim 50$ & 2030 & $0.47 \sim 0.61$
\\
37 & 245 & $1.18(3)$ & $0.889(12)$ & 0.91 &  43& $35 \sim 53$ & 2150 & $0.51 \sim 0.63$
\\
36 & 252 & $1.38(2)$ & $0.948(8)$ & 0.09 &  36& $29 \sim 46$ & 1744 & $0.46 \sim 0.58$
\\
35 & 259 & $1.26(2)$ & $0.931(7)$ & 1.15 &  40& $31 \sim 51$ & 2244 & $0.48 \sim 0.61$\\
\hline
\end{tabular}
\end{table*}

In \Fig{correlator-37}, the $0^{++}$ glueball correlator $G(t)/G(0)$ for
$\Nsmear  = 40$ at  $T=245$ MeV  is shown.   The statistical  errors are
estimated with the  jackknife analysis. The solid line  denotes the best
single hyperbolic cosine fitting as
$
	G(t)/G(0)
=
	C(e^{-m_{\rm G}t a_t} + e^{-m_{\rm G} (N_t - t) a_t})$,
performed in  the interval indicated  by vertical dotted lines.   In the
most   suitable  smearing   number  $N_{\rm   smear}^{\rm   best}$,  the
ground-state overlap $C$  is maximized and the mass  $m_G$ is minimized,
which  indicates the  achievement of  the ground-state  enhancement.  In
practical calculations,  the maximum  overlap and the  mass minimization
are achieved at  almost the same $N_{\rm smear}$, and  both of these two
conditions  would work  as  an indication  of  the maximal  ground-state
enhancement.  Here, we take the maximum ground-state overlap condition.

In  \Table{table}, we  summarize the  SU(3)  lattice QCD  result of  the
lowest $0^{++}$  glueball mass  $m_{\rm G}(T)$ at  various temperatures.
We  observe that  about  20 \%  mass  reduction of  the lowest  $0^{++}$
glueball near  $T_c$ as $m_{\rm G}(T) =  1.25 \pm 0.1$ GeV  for $0.8 T_c
\le T  \le T_c$ in comparison  with $m_{\rm G}(T\sim 0)  \simeq 1.5 \sim
1.7$ GeV \cite{morningstar,weingarten}.   To estimate the glueball size,
we search $N^{\rm best}_{\rm  smear}$ which achieves the maximum overlap
as $C = C^{\rm max}$. Considering  about 0.5 \% errors of $C$, we define
the ``smearing window'' $N^{\rm  window}_{\rm smear}$ where $C \ge 0.995
C^{\rm  max}$ is  satisfied.   From \Eq{size}  with $N_{\rm  smear}^{\rm
window}$, we  estimate the glueball size  as $R \simeq 0.5  \sim 0.6$ fm
both at low  and high temperatures.  In \Table{table},  we summarize the
lowest $0^{++}$  glueball mass $m_{\rm G}(T)$,  the ground-state overlap
$C^{\rm max}$,  uncorrelated $\chi^2/\Ndf$, $N^{\rm  best}_{\rm smear}$,
$N^{\rm window}_{\rm smear}$, the number of gauge configurations $N_{\rm
conf}$ and the glueball size $R$.
\section{SUMMARY}
We have  studied the $0^{++}$ glueball properties  at finite temperature
using SU(3) anisotropic quenched lattice  QCD with more than 1,000 gauge
configurations  at  each  temperature.   From the  temporal  correlation
analysis with  the smearing  method, we have  observed about 20  \% mass
reduction of the lowest $0^{++}$ glueball as $m_G(T) = 1.25 \pm 0.1$ GeV
for $0.8  T_c < T <T_c$,  while no significant change  has been observed
for the glueball size as $R \simeq 0.5 \sim 0.6$ fm.

\end{document}